\documentclass[a4paper]{elsarticle}

\usepackage{hyperref}
\usepackage{epsfig}

\def\fermi{{\it Fermi}-LAT }

\def\deg{$^{\circ}$}
\def\Cen~A{Centaurus~A}
\def\ph{\rm ph cm$^{-2}$ s$^{-1}$}

\begin{document}

\begin{frontmatter}

\title{Radio Galaxies with the Cherenkov Telescope Array}

\author[1,2,3]{R. Angioni\fnref{fn1}}

\author[1]{P. Grandi\corref{cor1}}
\ead{grandi@iasfbo.inaf.it}

\author[1,2]{E. Torresi}
\author[2,4]{C. Vignali}
\author[5]{J. ~Kn{\"o}dlseder}

\cortext[cor1]{Corresponding author}
\fntext[fn1]{Member of the International Max Planck Research School (IMPRS) for Astronomy and Astrophysics at the Universities of Bonn and Cologne}
\address[1]{INAF-IASFBO, Via Gobetti 101, I-40126 Bologna, Italy}
\address[2]{Dipartimento di Fisica e Astronomia, Universit\`{a} degli Studi di Bologna, viale Berti Pichat 6/2, I-40127 Bologna, Italy}
\address[3]{Max-Planck-Institut f\"ur Radioastronomie, Auf dem H\"ugel 69, D-53121 Bonn, Germany}
\address[4]{INAF--Osservatorio Astronomico di Bologna, Via Ranzani 1, I-40127 Bologna, Italy}
\address[5]{Institut de Recherche en Astrophysique et Plan\'etologie, 9 avenue Colonel-Roche, F-31028 Toulouse, Cedex 4, France}

\begin{abstract}
Misaligned AGN (MAGNs), i.e., radio-loud AGNs with the jet not pointing directly towards us, represent a new class of GeV emitters revealed by the \fermi\ space telescope. Although they comprise only a small fraction of the high-energy sources, MAGNs are extremely  interesting objects offering a different perspective to study high-energy processes with respect to blazars. The aim of this work is to evaluate the impact of the new-generation Cherenkov Telescope Array (CTA) on the MAGN class and propose possible observational strategies to optimize their detection. 
\end{abstract}

\begin{keyword}
\texttt{Galaxies: active; Galaxies: nuclei; Galaxies: jets; Gamma rays: galaxies}
\end{keyword}

\end{frontmatter}


\section{Introduction}
Radio-Loud Active Galactic Nuclei (RL AGNs)  constitute the majority of the extragalactic sources observed at Very High Energies (VHE, $E>100$~GeV; see, for example, the TeVCat online catalog\footnote{URL: http://tevcat.uchicago.edu/}).
RL AGNs are characterized by the presence of well-collimated jets of relativistic plasma ejected from the nuclear region \cite{miley80}. The triggering and launching mechanism of these jets, and therefore the physical conditions and parameters under which an AGN becomes radio-loud, are one of the most debated and studied topics in this field \cite{bla}.

Jets host a population of relativistic particles which emit
non-thermal radiation across the whole electromagnetic spectrum. The Spectral
Energy Distribution (SED) associated with this emission
has a characteristic shape comprising two peaks: one at low energy (radio
to optical/soft X-rays) and one at high energy (hard X-rays to TeV) \cite{fo98} .
The low-energy peak is confidently associated with synchrotron
emission from a population of relativistic electrons in the jet, while
the high-energy emission process is not well understood. The most
popular model invokes Inverse Compton (IC) up-scattering of low-energy seed photons by the
same relativistic electrons responsible for the synchrotron radiation.
The seed photons can be provided by the electron synchrotron emission
itself, in which case the process is referred to as Synchrotron Self-Compton (SSC) \cite{ma92}, or by an external photon field (such as the accretion disk,
the Broad Line Region, the Narrow Line Region, the torus,  or the CMB radiation), referred to as External
Compton (EC) process \cite{sik,tav}. 
These models are collectively called
\textit{leptonic} models since all the observed emission is ascribed
to relativistic electrons. In recent years, it has become clear that a
leptonic modeling of the SED of RL AGNs is not capable of reproducing
all the observed properties. Therefore, alternative models have been
developed which consider, alongside the relativistic leptons, a
population of relativistic protons. This component can give the main
contribution to the observed emission, particularly at the highest
energies, reproducing the flat VHE spectra that are observed in some
sources, while the electron IC component provides the hard X-ray/soft
$\gamma$-ray emission. The main emission processes involved are proton
synchrotron emission, photo-pion production, and cascade emission from
secondary particles produced by the interactions of protons with
ambient photons or themselves. These models are called
\textit{hadronic} or \textit{lepto-hadronic} models; see, e.g.,
\cite{bott10}, \cite{bott13} for a review on leptonic and hadronic models.

If the jet of a radio-loud AGN is closely aligned with the observer's
line of sight (l.o.s), its radiation is strongly beamed and amplified
via relativistic Doppler effects. In this case, the AGN is called  a
blazar, and the non-thermal jet emission typically dominates over other contributions to the observed SED. 
The blazar class includes Flat Spectrum Radio Quasars (FSRQs) and BL Lacertae (BL Lacs). 
FSRQs are powerful sources displaying strong broad optical emission lines, while BL
Lacs are less luminous than FSRQs and lack emission lines (down to equivalent widths of a few \AA) in their optical spectra.

In the framework of the
orientation-based unified model \cite{bar89,urry95} radio galaxies
(and Steep Spectrum Radio Quasars, SSRQs) are the radio-loud AGNs with the jet pointed away from the observer 
and are therefore referred to as
Misaligned AGNs. Radio galaxies can be divided into two classes:
FR~I at low radio powers, which present decelerating jets and edge-darkened
diffuse lobes, and FR~II at high radio powers, with large-scale relativistic
jets and edge-brightened lobes \cite{fan74}. The transition between these two classes 
happens  at a luminosity of $\sim10^{25}$ W~Hz$^{-1}$~sr$^{-1}$ at 178~MHz. 
FR~I and FR~II radio galaxies are considered the parent population  of  BL
Lacs and FSRQs, respectively  (\cite{p190, p290,p91,p92}). 
Recently, the new class
of ``FR~0'' radio galaxies has been proposed
(e.g., \cite{bal15} and references therein). These sources share similar nuclear and host properties with FR~Is, but lack extended radio emission. 
Though their behavior is still poorly understood, they  appear to
represent the majority of the population of local RL AGNs.

Because of
the larger inclination angle between the jet and our l.o.s. with respect to blazars,  the observed
non-thermal emission from radio galaxies is not significantly
Doppler-boosted, therefore these sources  have a less
jet-dominated SED. Hence, radio galaxies provide a view of AGN
jets which is less biased by relativistic effects, allowing us to observe both the
jets and the accretion process and potentially establish a 
connection between the two. This represents the first, fundamental step to start a proper investigation of the origin of radio-loudness in general \cite{mar02,cha09,cha11}.
Moreover, radio galaxies 
allow us to investigate the presence of a transverse jet structure,
which has been proposed to explain the observed SED of FR~Is.
Structured (multi-zone) jets are indeed promising in describing the high-energy radiation of radio galaxies.
SSC models, when applied to RGs, require (modest) beaming factors ($\delta \sim 2-3$)\footnote{$\delta=[\Gamma(1-\beta cos(\theta))]^{-1}$, being $\theta$ the inclination angle and $\Gamma=(1-\beta^2)^{-1/2}$ the Lorentz factor and $\beta$ the bulk jet velocities in unit of the speed of light.} implying, for large inclination angles,  small Lorentz factors 
($\Gamma\sim 2-3$) \cite{chi1,abdo9a,abdo9b,abdo10a,mi}. Slow jets are in conflicts with the idea that RGs are the parent population of blazars. Very high jet velocities ($\Gamma \sim 10$ or more) are indeed found in AGN with jet pointed directly to the observer \cite{ghi15}.

If the condition of a homogeneously emitting region as assumed in a SSC one-zone model is relaxed, 
an efficient (radiative) feedback between different
regions can explain the observed IC peak of radio galaxies (without violating the unified models \cite{mi}). 
This can be explained by a decelerating flow \cite{mar} and a spine-layer jet \cite{mar, gg2005, tave08, tave14}.
In one case, the presence of regions at different velocities along the relativistic flow is assumed;
in the other case, the flow is supposed to be fast in the inner part (the region observed in blazars) and slow in the external envelope. 
Interestingly enough, a limb-brightened structure of the jet has been discovered 
in Mrk~501 \cite{giro2004}, in M~87 \cite{ko2007,na2014} and in NGC~1275 \cite{tave14}, thus supporting the latter scenario.

The weaker Doppler boosting of MAGN emission is particularly evident
in the $\gamma$-ray band, where blazars represent the great majority of the 
observed AGNs. For example, in the latest \fermi\ AGN
catalog, the 3LAC \cite{3LAC}, MAGNs with a solid identification
constitute only $\sim$1-2\% of all sources.
The second \fermi\ catalog of
hard spectrum sources (2FHL \cite{2FHL}), which covers the 
energy range between 50 and 2000~GeV, includes 271 sources
associated with AGNs, of which 6 are radio galaxies (2\%  of the entire sample). 
All the  2FHL radio galaxies  have been detected by Cherenkov Telescopes with the only exception of 3C~264 that, however, has no detection  above ~170~GeV in the \fermi\ band.
The TeV MAGNs  are well known local FR I radio galaxies
(see the TeVCat online catalog), i.e. 
NGC~1275 \cite{ale2012}, M~87 \cite{aha03}, and Centaurus~A
\cite{aha09}, plus the transitional FR~I-BL Lac source IC 310
\cite{aleski09}. Recently, the detection of the FR~I
radio galaxy PKS~0625$-$35 was announced by the H.E.S.S. collaboration 
\cite{dyr15}. 

Although TeV observations of MAGNs have played an important role in
investigating the open questions in high-energy studies of jets, the limited number of detected sources, 
along with the low signal-to-noise spectra produced by the current Imaging
Atmospheric Cherenkov Telescopes (IACT), i.e. MAGIC, H.E.S.S.
and VERITAS, do not allow us to draw general conclusions. 

The new-generation facility for VHE astronomy, the Cherenkov
  Telescope Array (CTA) \cite{ach13}, is expected to achieve
order-of-magnitude improvements in sensitivity and energy range with
respect to previous facilities which operate in the same energy domain.
These capabilities will be achieved through the deployment of a large
 number of Cherenkov telescopes of different sizes at  two
sites (Northern and Southern hemisphere) for full sky coverage. In the final configuration,  the
Northern array will include 4 Large Size Telescopes (LST, 23m diameter) and 15
Medium Size Telescopes (MST, 12m diameter). The Southern array will include 4
LSTs, 25 MSTs, and 70 Small Size Telescopes (SST, 4m diameter). The three
classes are designed to cover the low ($20-200$~GeV), intermediate ($0.1-10$~TeV), and high-energy
ranges (up to 300~TeV), respectively. Because of the different array configurations,
the Southern array will have a better sensitivity, especially at
energies $>$ 1~TeV.

Thanks to these remarkable improvements, the CTA is also expected to
increase our ability in revealing  MAGNs in the TeV sky. Our aim is
therefore to evaluate its impact on VHE studies of non-blazar RL AGNs
starting from the exploration of the MAGN sample revealed in the GeV
band by {\it Fermi}.

In Section~2 we describe the sample of radio galaxies
observed in the GeV band, which is the starting point for our study.
In Section~3 we describe the performed simulations and the results. In
Section~4 we present more general simulations that allow us
to estimate the chances of a CTA detection for any \fermi\ AGN, given
its spectral parameters in the 1-100~GeV band. In Section~5 we draw the conclusions
and discuss the optimal observing strategy to detect more MAGNs,
in light of our results.

\section{The \fermi\ radio galaxy sample} 

The third  \fermi\  AGN catalog (3LAC) \cite{3LAC}, based on 4 years of observations, provides the most up-to-date list of identified AGNs emitting in the GeV band.  In our sample, we include all the radio galaxies with a solid 
counterpart reported in the 3LAC. We also
add the radio galaxy  3C~120, firmly established as a $\gamma$-ray
source by \cite{abdo10b,cas15}  and  the FR~0 radio galaxy
Tol 1326$-$379, which has recently been associated with the Fermi source 3FGL J1330.0$-$3818 \cite{gra16}. We decided to consider only the
sources with redshift $z<0.15$ to avoid any significant $\gamma\gamma$ absorption  by the Extragalactic Background Light (EBL).
This excludes all the SSRQs in the 3LAC catalog from our MAGN sample, leaving 17 radio galaxies. 
 
Table~\ref{tab3lac} reports the $\gamma$-ray properties of the 17 \fermi\ radio galaxies. 
They  are weak sources with 1--100~GeV fluxes of  the order of $\sim 10^{-10} - 10^{-8}$~photons~cm$^{-2}$~s$^{-1}$, and have power-law spectral indices ($\Gamma_{\mathrm{Fermi}}$) in the range 1.8--2.8. In only two cases, NGC~6251 and NGC~1275, the spectra are
curved and better reproduced by a logparabola model, rather than a power-law \cite{3LAC,3FGL}.  In the logparabola model\footnote{$F(E)=k E^{[\Gamma_{\mathrm{Fermi}} -\beta(\log(E)]}$}  the deviation from the power law is modeled by the $\beta$ parameter, also listed in Table~\ref{tab3lac}.

In Table~\ref{tevrg}, TeV spectral slopes and fluxes  ($>100$~GeV)  of the five sources detected also by Cherenkov Telescopes up to 2$-$10~TeV are listed together with  their 2FHL properties.
In spite of the partial overlap of the energy bands, the \fermi\ and IACT results are consistent within the large uncertainties. 
\begin{table}[h!tbp]
\footnotesize{
\caption{\fermi\ radio galaxies}
\begin{tabular}{|l|l|l|l|l|l|l|}
\hline
 \multicolumn{1}{|c|}{3FGL Name} &
  \multicolumn{1}{|c|}{Object} &
  \multicolumn{1}{|c|}{Class} &
  \multicolumn{1}{|c|}{z} &
  \multicolumn{1}{|c|}{Model$^a$} &
  \multicolumn{1}{|c|}{$\Gamma_{\mathrm{Fermi}}$} &

  \multicolumn{1}{|c|}{F$_{\mathrm{1-100~GeV}}$} \\
  \multicolumn{1}{|c|}{} &
  \multicolumn{1}{|c|}{} &
  \multicolumn{1}{|c|}{} &
  \multicolumn{1}{|c|}{} &
  \multicolumn{1}{|c|}{} &
  \multicolumn{1}{|c|}{$\beta$} &
  \multicolumn{1}{|c|}{phot cm$^{-2}$ s$^{-1}$} \\
\hline
 3FGL J0308.6+0408   & 3C~78 & FRI & 0.029 & PL & 2.1$\pm$0.1 &  (5.9$\pm$1.0)$\times10^{-10}$\\
3FGL J0316.6+4119    &  IC~310 & FRI & 0.018 & PL & 1.9$\pm$0.1 & (6.6$\pm$1.5)$\times10^{-10}$\\
3FGL J0319.8+4130	  &NGC~1275 & FRI & 0.018 & LogPar & 1.98$\pm$0.01 & (2.12$\pm$0.04)$\times10^{-8}$\\
                                    &                   &        &          &              &  (6.5$\pm$0.7)$\times10^{-2}$& \\
  3FGL J0322.5$-$3721& Fornax~A$^{b}$ & FRI & 0.0058 & PL & 2.2$\pm$0.1 &  (4.7$\pm$0.8)$\times10^{-10}$\\
  3FGL J0334.2+3915&B2~0331+39 & FRI & 0.0206 & PL & 2.1$\pm$0.2 &  (3.7$\pm$0.9)$\times10^{-10}$\\
  3FGL J0418.5+3813c	&3C~111 & FRII & 0.049 & PL & 2.79$\pm$0.08 &  (7.3$\pm$1.3)$\times10^{-10}$\\
 &  3C~120$^c$ & FRI & 0.033 & PL & 2.7$\pm$0.1 &  (5.4$\pm$1.1)$\times10^{-10}$\\
3FGL J0519.2$-$4542&  Pictor~A & FRII & 0.035 & PL & 2.5$\pm$0.2 &  (3.2$\pm$0.7)$\times10^{-10}$\\
  3FGL J0627.0$-$3529& PKS~0625$-$35 & FRI & 0.055 & PL & 1.87$\pm$0.06 &  (1.4$\pm$0.1)$\times10^{-9}$\\
3FGL J0758.7+3747&  3C~189 & FRI & 0.0428 & PL & 2.2$\pm$0.2 &  (2.5$\pm$0.7)$\times10^{-10}$\\
 3FGL J1145.1+1935&  3C~264 & FRI & 0.0217 & PL & 2.0$\pm$0.2 &  (2.7$\pm$0.7)$\times10^{-10}$\\
  3FGL J1230.9+1224	& M~87 & FRI & 0.0042 & PL & 2.04$\pm$0.07 &  (1.3$\pm$0.1)$\times10^{-9}$\\
 3FGL J1325.4$-$4301& Cen~A core& FRI & 0.0018 & PL & 2.70$\pm$0.02 &  (3.4$\pm$0.2)$\times10^{-9}$\\
 3FGL J1330.0$-$3818& Tol~1326$-$379 & FR0 & 0.0284 & PL & 2.8$\pm0.1$ &  (3.1$\pm$0.8)$\times10^{-10}$\\
 3FGL J1346.6$-$6027	& Cen~B & FRI & 0.013 & PL & 2.32$\pm$0.01 &  (2.0$\pm$0.2)$\times10^{-9}$\\
 3FGL J1442.6+5156& 3C~303 & FRII & 0.141 & PL & 1.9$\pm$0.2 &  (1.9$\pm$0.5)$\times10^{-10}$\\
 3FGL J1630.6+8232& NGC~6251 & FRI & 0.025 & LogPar& 2.04$\pm$0.08 & (1.3$\pm$0.1)$\times10^{-10}$\\
       &               &       &            &            &0.17$\pm0.05$ &\\
\hline
\multicolumn{7}{l}{$^a$ -- PL - $F(E)=kE^{-\Gamma_{\mathrm Fermi}}$; LogPar - $F(E)=kE^{-[\Gamma_{\mathrm Fermi} + \beta \log(E)]}$}. \\
\multicolumn{7}{l}{$^b$ -- $\gamma$-ray emission associated to radio lobes \cite{fornax}}. \\
\multicolumn{7}{l}{$^c$ -- Spectral parameters from \cite{cas15}}. \\
\end{tabular}
\label{tab3lac}}
\end{table}

\begin{table}[h!]
\footnotesize{
\caption{Sub-sample of MAGNs with TeV detection. All the data are fitted with a power law model. For variable sources,   
minimum and maximum fluxes (and  relative spectral slopes)  measured by the Cherenkov telescopes are reported.}
\begin{tabular}{|l|l|l|l|l|l|l|}
\hline
  \multicolumn{1}{|c|}{Object$^{a}$}&
  \multicolumn{1}{|c|}{$\Gamma_{\mathrm{\small 2FHL}}$} &
  \multicolumn{1}{|c|}{F$_{\mathrm{2FHL}}^b$ }&
  \multicolumn{1}{|l|}{$\Gamma_{\mathrm{TeV}}$} &
  \multicolumn{1}{|c|}{F$_{\mathrm{TeV}}^b$}&
   \multicolumn{1}{|c|}{TeV}\\ 
   \multicolumn{1}{|c|}{} &
  \multicolumn{1}{|c|}{$>50$ GeV } &
  \multicolumn{1}{c|}{} &
  \multicolumn{1}{|c|}{} &
  \multicolumn{1}{|c|}{ $>100$ GeV$^c$}&
  \multicolumn{1}{|c|}{ref} \\
 \hline
IC~310$^{d}$    & 1.3$\pm0.4$ & 22$\pm$09 & 1.81-1.85 & 6-43& \cite{ale14ic}\\
NGC~1275 & 3.0$\pm0.5$ & 51$\pm$13& 4.1$\pm0.7_{\mathrm{stat}}\pm0.3_{\mathrm{sys}}$ & 13$\pm2_{\mathrm{stat}}\pm3_{\mathrm{sys}}$&\cite{ale2012}\\
PKS~0625$-$35 & 1.9$\pm0.5$ & 29$\pm$11 & 2.8$\pm0.5$ & 20$^e$&\cite{dyr15}\\
M~87$^{d}$ & 2.3$\pm0.6$ & 22$\pm$10& 2.2-2.6 & 30-60&\cite{aliu12}\\
Cen~A & 2.6$\pm0.8$ & 18$\pm9$& 2.73$\pm0.45_{\mathrm{stat}}\pm0.2_{\mathrm{sys}}$ & 7.6$^e$&\cite{aha09}\\
\hline
\end{tabular}
$^{a}$ -- The 2FHL catalog reports the detection of 3C~264. It was not included in Table because its \fermi spectrum does not extend above 171~GeV \cite{2FHL}.\\
$^b$ -- In unit of $10^{-12}$~photon cm$^{-2}$ s$^{-1}$.\\ 
$^c$ -- Fluxes provided by different Cherenkov telescopes can cover different energy bands. $F_{\mathrm{TeV}}$ was extrapolated down to 100 GeV, when necessary.\\
$^d$ -- Variable source.\\
$^e$ -- Flux uncertainty of the order of $~\sim30\%$.
\label{tevrg}
}
\end{table}

Finally, we note that our \fermi\ sample is obviously  biased towards sources with a
high-energy (approximately in the MeV-GeV band) SED peak.  It is possible that we are missing MAGN  peaking at very high energies. They could be  below the  \fermi\  sensitivity threshold and 
emerge in the CTA band.  IC~310 is an example, although less extreme. It is barely detected with a very low flux by \fermi but firmly detected with a flat spectrum at VHE \cite{nero10,ale14ic} with no sign of a falling trend up to 10~TeV. 

\section{CTA simulations of \fermi\ radio galaxies}
The connection between the \fermi\  and IACT data is complex.
First of all, data in these two bands are usually not contemporaneous. 
This is relevant because MAGNs can vary both
in the LAT band \cite{ale14n} and in the TeV band
\cite{aha06,acc08,aliu12,abr12, ale14ic} and the variability is not necessarily correlated.
Moreover,  while an extrapolation of the \fermi\ spectrum fits well the non-simultaneous  MAGIC TeV data points of M~87  in a low state  \cite{ale12l}, 
a curvature in the overall spectrum of NGC~1275 is clearly established
by two simultaneous  MAGIC - \fermi\ campaigns \cite{ale14n}.
In addition,  the MeV-GeV and the TeV emissions might be produced by distinct components as in the case of Centaurus~A, where a second spectral component seems to emerge between the \fermi\ and H.E.S.S. spectra \cite{sa13, fra14}. 
This spectral component could be the signature of efficient pulsar-like electron acceleration mechanisms occurring in the black-hole magnetosphere \cite{ri09} or could mark the presence 
of an hadronic process \cite{pe14}; alternatively, it could be related to a population of millisecond pulsars.  Another, even more exciting, possibility is that such a component is produced by heavy dark matter (DM) particles clustered around the black hole \cite{bro}.
The discovery of the VHE hardening of the Centaurus~A spectrum opens new interesting scenarios that CTA will be able to explore.

Finally we note that a GeV contribution from the radio lobes has been revealed in  two nearby sources: CenA \cite{lobi} and Fornax~A \cite{fornax}.  
The poor LAT spatial resolution ($\sim 0.2^{\circ}$ at 10 GeV and larger at lower energies)  does not allow us to disentangle different emission regions in other MAGN. However GeV flux variability   observed in  3C~120 \cite{cas15, jan16},  3C~111 \cite{3c111}, NGC~1275 \cite{kat, muk}, M~87 \cite{aliu12} and IC~310 \cite{ale14ic} suggest that most of the high and very high energy photons are  dissipated in compact region.\\

Keeping in mind all the caveats reported above, we decided to simulate the GeV-TeV spectra of radio galaxies assuming the compact jet core as
the main source of GeV photons. Studies of extended emitting regions  require further CTA simulations that are beyond the scope of this first explorative work on the CTA performances.  

We considered different spectral shapes.
At first, a simple extrapolation of the \fermi\ power-law (PL) was assumed, then an 
exponential cutoff\footnote{$F(E)=k (\frac{E}{300~GeV})^{-\Gamma_{\mathrm{Fermi}}} e^{(-E/E_{\mathrm{c}})}$ } was included. 
 We considered three possible cutoffs at decreasing energies $E_{\mathrm c}$=1~TeV, $E_{\mathrm c}$=500~GeV, and $E_{\mathrm c}$=100~GeV  to 
take into account different spectral steepenings.  As a spectral curvature is already present in the \fermi\ spectrum of NGC~6251, 
we extrapolated the logparabola model of the 3FGL catalog (rather than a power-law) and, in addition, we tested a power-law with a cutoff at 100~GeV as another possible parameterization of the 
high-energy steepness. Actually, it is usually difficult to distinguish between a logparabola model and a cutoff power-law with the current data, as also shown by the Fermi-MAGIC study of NGC~1275 \cite{ale14n}.
Obviously, our approach reflects the lack of  information on the SED of MAGNs in the TeV domain.
It is mainly driven by the \fermi\ and IACT observations of M~87 and NGC~1275. It is also evident that we are adopting the less favorable scenarios, not considering any flattening of the very high-energy spectrum, as observed in Centaurus~A. \\

\begin{figure*}[htbp]
\begin{center}
\includegraphics[width=6.0cm]{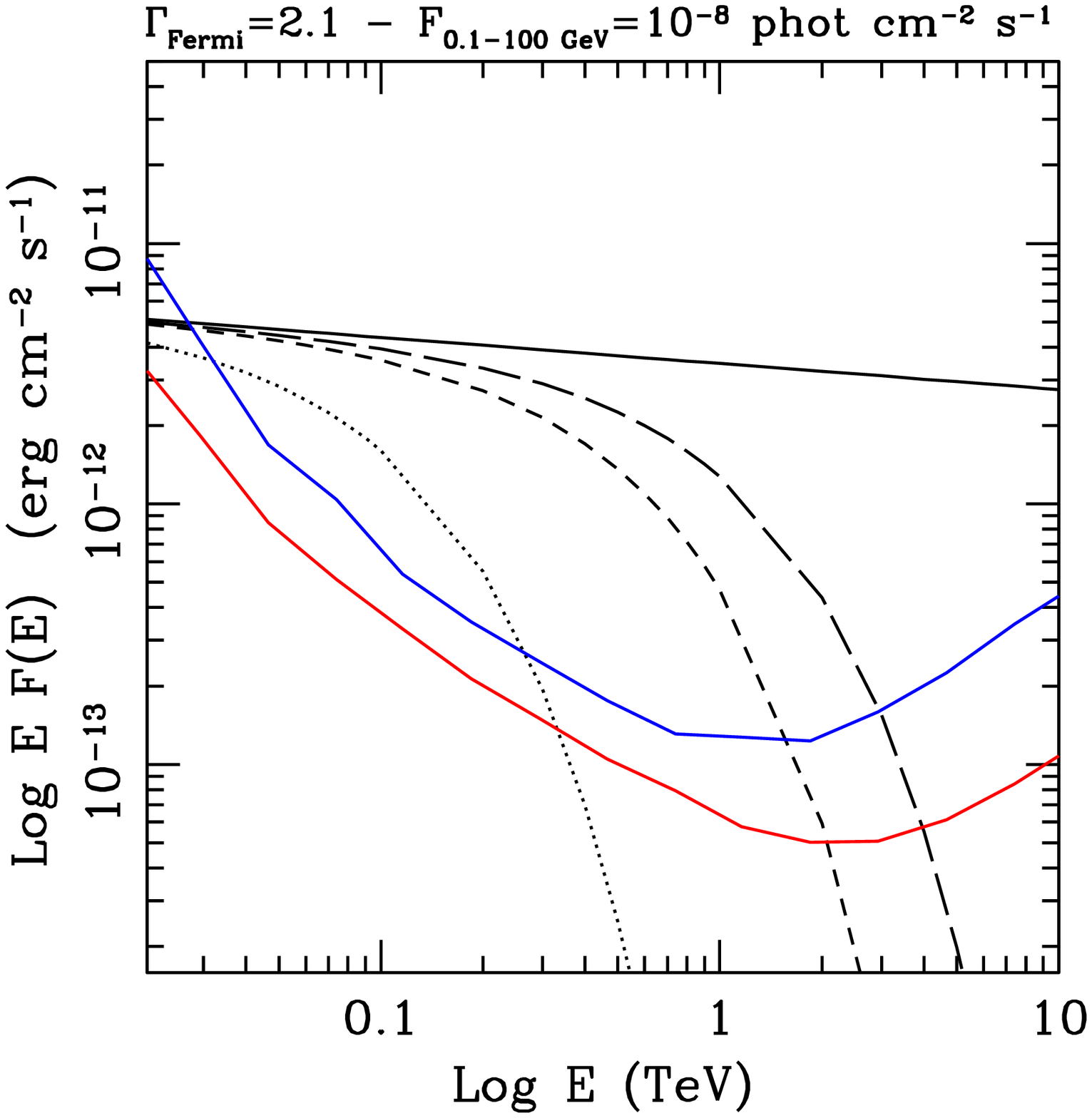}
\includegraphics[width=6.0cm]{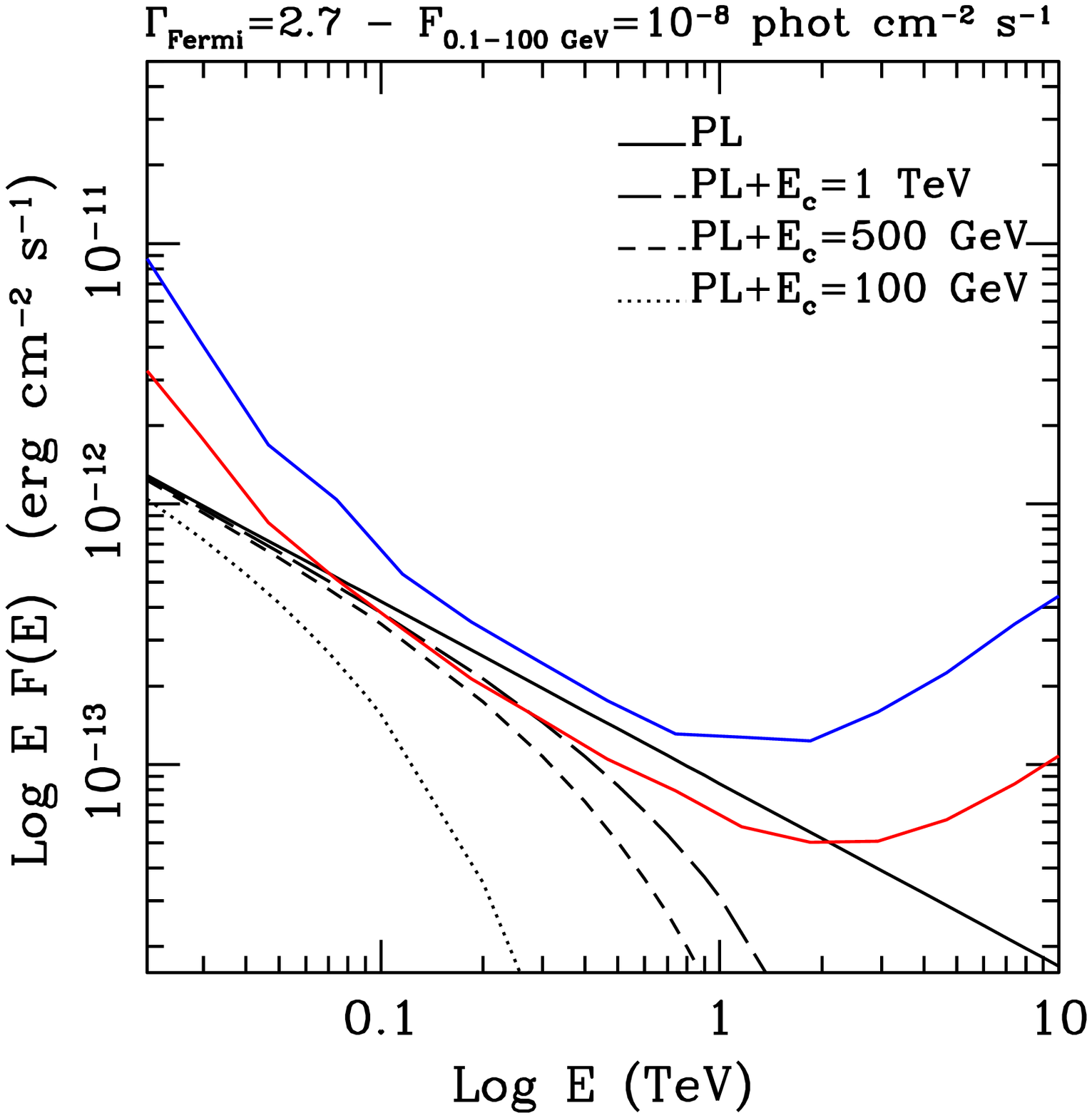}
\end{center}
\caption{Example of simulated CTA models (black curves, legend on the right panel),
  compared with differential sensitivity curves from CTA North (blue
  line) and CTA South (red line), for two different input  spectra. The relevance of the power-law slope to a possible CTA
  detection is evident, as well as the effects of different
  exponential cutoff energies.}
\label{Tsvsg}
\end{figure*}

In Fig.~\ref{Tsvsg}  the adopted models are shown along with the
current differential sensitivity curves  for 50 hours of observation for the
Northern (blue line) and the Southern CTA array (red line). 
As an example, we plot the  simulated models for two different power-law spectral slopes, i.e  $\Gamma_{\mathrm{Fermi}}=$2.1 {\it (left panel)} and  
$\Gamma_{\mathrm{Fermi}}=$2.7 {\it (right panel)} and same 
input flux $F_{\mathrm{1-100~GeV}}=10^{-8}$ ph {\rm cm$^{-2}$ s$^{-1}$}.
It is clear from  Fig.~\ref{Tsvsg} that, apart from the cutoff energy position ($E_{\mathrm{}}$),  the spectral index is a fundamental  parameter: flat \fermi\ sources are more suitable CTA targets. This will be further confirmed by the specific simulations of the MAGN sample discussed later in detail. 
Finally, we note that, as expected, the best performance is provided by the Southern Array which consists of a larger number of telescopes.\\

The spectra of the radio galaxies in Table~\ref{tab3lac} were simulated in the energy range 0.02--100~TeV using the software \textit{ctools v1.0}\footnote{http://cta.irap.omp.eu/ctools} \cite{kno16}, developed for the scientific analysis of CTA data. Also Fornax~A, that is part of the sample, was considered, although the $\gamma$$-$ray emission has been recently associated with the radio lobes \cite{fornax}. As our simulations are based on point-like sources, the results should be taken with caution. 
We did not consider the sources with TeV observations,  with the exception of PKS~0625$-$35 for which only a recent claim of detection has been reported \cite{dyr15}.

For each source,  an event list corresponding to a particular model was produced considering a 5$^{\circ}$ circular region of interest (ROI) centered on the point-like target. 
The most recent Instrument Response Functions (IRF, version from 2015$-$05$-$05)\footnote{https://www.cta-observatory.org/science/cta-performance/} provided by the CTA, supplying information on effective area, point spread function, energy dispersion  and instrumental background of the CTA configurations in the two hemispheres were assumed for an observation time of 50 hours. 

A standard unbinned likelihood analysis was then performed to test the significance of the source. 
We considered the Test Statistic that is defined  as TS = 2[logL$_{\mathrm{s}}$-logL$_{\mathrm{0}}$] where L$_s$ is the maximum likelihood value for a model with our source at a specified location and L$_0$ is the maximum likelihood value for a model without the  source \cite{mat96}. Only the normalization and the spectral index  were allowed to vary.
With the square root of the TS corresponding approximately to the detection significance, we consider thresholds of TS$>$100 (corresponding to approximately  10$\sigma$) and TS$>$25 ($\sim5\sigma$) to assess the detectability of MAGNs by CTA.

 \section{Results}
 
The results of the simulations are summarized in  Table~\ref{tabpl}.
We expect that 7 out of 12 MAGNs (8 out 13 if Fornax~A is also considered) will be detected at Very High Energies if the simulated CTA
spectrum is a direct extrapolation of the \fermi\ power-law. 
The number of likely CTA candidates obviously decreases with the
steepening of the spectrum in the TeV band. However, half of the
sources are still above the CTA sensitivity threshold  (and three at a
significance level larger than $\sim10\sigma$)  if  the spectral cutoff
occurs at energies $\ge$ 500~GeV.   

\begin{figure*}[!htbp]
\begin{center}

\includegraphics[width=10.cm]{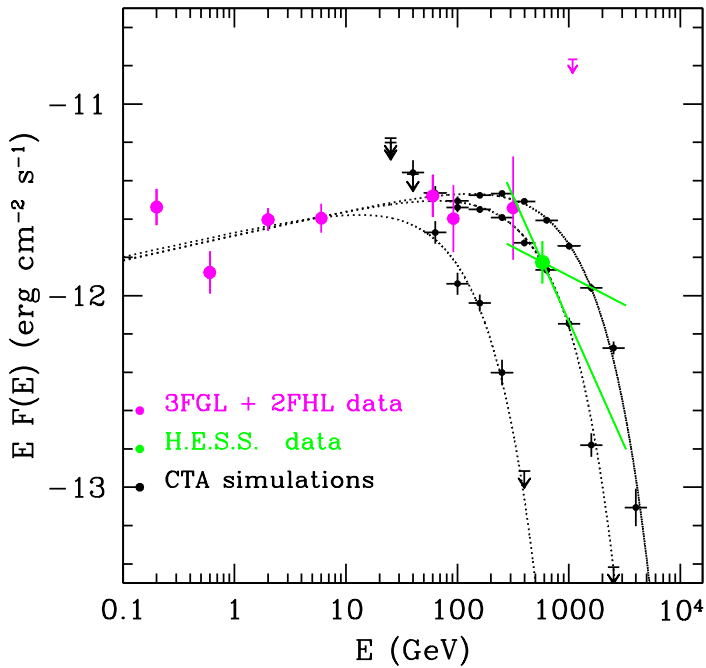}
\end{center}
\caption{Comparison between observed  and simulated data for PKS
  0625-35. \fermi\ data (pink points) and H.E.S.S. best-fit spectrum (green point
  and lines) are plotted along with our simulated spectra (black points and
  curves), corresponding to three different input models, i.e. a power-law with a cutoff energy at 100~GeV, 500~GeV and 1~TeV.
   \fermi\ data are from the 3FGL
  \cite{3LAC} and the 2FHL \cite{2FHL} catalogs. H.E.S.S. data are
  from \cite{dyr15}.}
\label{pks}
\end{figure*}

PKS~0625$-$35 is the only MAGN with  a TS value larger than 100 for each
tested model, and indeed a TeV detection has been recently reported by
the H.E.S.S. collaboration \cite{dyr15}.
In Figure~\ref{pks}, the  CTA simulated spectra of PKS~0625$-$35 are shown together with  the \fermi\ and H.E.S.S. data. 
The H.E.S.S. data (falling between two CTA models) suggest a high-energy cutoff between 0.5 and 1 TeV.
In order to quantify the CTA ability to discriminate among different models, the simulation of PKS~0625$-$35 with $E_{\mathrm cut}=500$ GeV was fitted with both a power law and a cutoff power law.  All the parameters (spectral slope, normalization, and $E_{\mathrm cut}$) were freely adjusted by the fit.
A likelihood ratio test TScurve=2[log L(Power Law$+$cutoff) - log L(Power Law)]  was then calculated.
As TScurve is distributed as $\chi^2$  with 1 degree of freedom \cite{nolan}, we can assume  that a curved spectrum  is better than a power law  when TScurve is  larger than 16 (corresponding to $\sim 4 \sigma$ significance for the curvature). The TScurve value of $\sim$ 400 obtained for PKS~0625$-$35 shows that a spectral bending is statistically preferred to a power law and  
attests that 
the CTA Southern array will not only be able to observe this source  but also to discriminate among different spectral shapes. 
Finally,  we note that PKS~0625$-$35 could be detected in only 5 hours with a TS $>$ 600 for $E_{\mathrm{cut}} \ge 500$~GeV.

Faint ($F_{\mathrm{1-100~GeV}} < 10^{-10}$ phot cm$^{-2}$ s$^{-1}$) and steep ($\Gamma_{\mathrm{Fermi}} \ge 2.5$) MAGNs have TS values below the threshold of 25 (see Table 3), independently of the adopted input model (see also Section~2).  We note, however, that radio galaxies are variable sources.
For example,  3C~120 underwent several flares, reaching in some cases  fluxes of $F_{\mathrm{> 100~MeV}}\sim 10^{-7}-10^{-6}$ ph~ cm$^{-2}$~s$^{-1}$ \cite{jan16}, 
and 3C~111 was detected in a very high state \footnote{https://www.bu.edu/blazars/VLBA$\_$GLAST/3c111.html} in more than one observation,  exceeding the flux of $10^{-7}$ ph {\rm cm$^{-2}$ s$^{-1}$}. If the jet perturbations responsible for the \fermi\ flares also cause bursts at TeV energies,  3C~120 and 3C~111 could be detected by the CTA (assuming power-law spectra) during flaring episodes.

Finally, we observe that a  direct extrapolation of the logparabola model for NGC~6251 gives a non-detection with the CTA.  However, a less abrupt decrease of the VHE emission,  as described by a 
power-law with a cutoff at 100~GeV, provides a marginal detection of the source (TS $\sim$ 25).

 In summary, unless MAGN emission falls down rapidly at the extreme end of the LAT  band, the new generation of Cherenkov telescopes can provide
high signal-to-noise spectra (see Figure 2) for several misaligned AGN permitting the extension of their  Spectral Energy Distribution above 1 TeV.
This will allow a deeper investigation of the still debate nature of the second emission peak  and, in particular, to confirm or reject the presence of an additional high-energy component (which presently has been observed only in Cen~A).

 \label{sec:meth}

\begin{table*}[!htbp]
\caption{Results of the simulated CTA observations for our radio galaxy
sample. TS$^{a}$ is the statistical significance of the source for the different spectral models tested here.}
\small{
\begin{center}
\begin{tabular}{|l|lllll|}
\hline
 \multicolumn{1}{|l|}{Source} & $k_{\mathrm{PL}}^b$ &TS$_{\mathrm{PL}}$ & TS$_{\mathrm{1~{TeV}}}$ &TS$_{\mathrm{0.5~{TeV}}}$&TS$_{\mathrm{0.1~{TeV}}}$\\
 \hline

3C~78                  & 47    & 3113   & 256 & 96 & ...\\
Fornax~A$^c$     &  20      &1671 & 100 & 25 & ...\\
B2~0331+39        & 24    &1049      & 65 &  ... & ...\\
3C~111                &  1.6  &... & ... & ... & ...\\
3C~120                & 1.9    &... & ...& ... & ...\\
Pictor~A               &  3.3   & ... &... & ... & ...\\
PKS~0625$-$35    & 300   & 154480  & 16987 & 7294 & 287\\
3C~189              & 13    &  321   & ...  & ... &...\\
3C~264              &  33   &2129    & 128 &  44& ...\\
Tol~1326$-$379    & 0.7     & ... & ... & ... & ...\\
Cen~B               &  48    & 4066 & 520 & 212 & ...\\
3C~303               &  32    &3010    & 134 & 55 & ...\\
NGC~6251$^d$    & 55.3 &  &   &  & 25 \\
\hline
\end{tabular}
\end{center}}
$^a$ Resulting TS from simulations assuming a simple extrapolation
of the \fermi\ power-law, and an exponential cutoff at 1~TeV, 0.5~TeV and 0.1~TeV, respectively. (...) indicates TS$<25$. 
\\$^b$ Normalization in units of $10^{-19}$ photons cm$^{-2}$
s$^{-1}$ MeV$^{-1}$ at 300~GeV.
\\$^c$ Fornax A $\gamma$-ray emission is probably produced in the extended radio lobes. See text for details. 
\\$^d$ Only a power-law with cutoff at 100~GeV was simulated for NGC~6251, as a spectral bending  is already present  in the \fermi\ band.
\label{tabpl}
\end{table*}

\section{Extension of the MAGN results}
In order to generalize the results obtained for the \fermi\ radio galaxies, we simulated a grid of  possible CTA observations with the Northern and Southern arrays, separately with the aim of producing a diagnostic plot to verify the detectability of any AGN in the TeV band,  provided that its spectral slope and flux in the \fermi\ band are known.
The simulation of the event files and then the likelihood analysis are the same as described in the previous section. We explored  $\gamma$-ray targets with fluxes  (1,2,4,6, 8)$\times(10^{-7},10^{-8},10^{-9},10^{-10})$ \ph~ between 1 and 100~GeV (the flux ranges covered by the AGN in the 3LAC catalog) and power-law spectral slope values ranging from 1.8 to 4.0 with an incremental step of $\Delta\Gamma=0.1$.  We did not consider $\Gamma_{\mathrm{Fermi}} < 1.8$ because, as indicated by the MAGN simulations, hard sources, even characterized by a LAT flux as low as  $10^{-10}$  \ph, can easily reach a  TS larger than 100. Both spectral slopes and normalizations were allowed to vary during the likelihood analysis. 

\begin{figure*}[htbp]
\includegraphics[width=10cm]{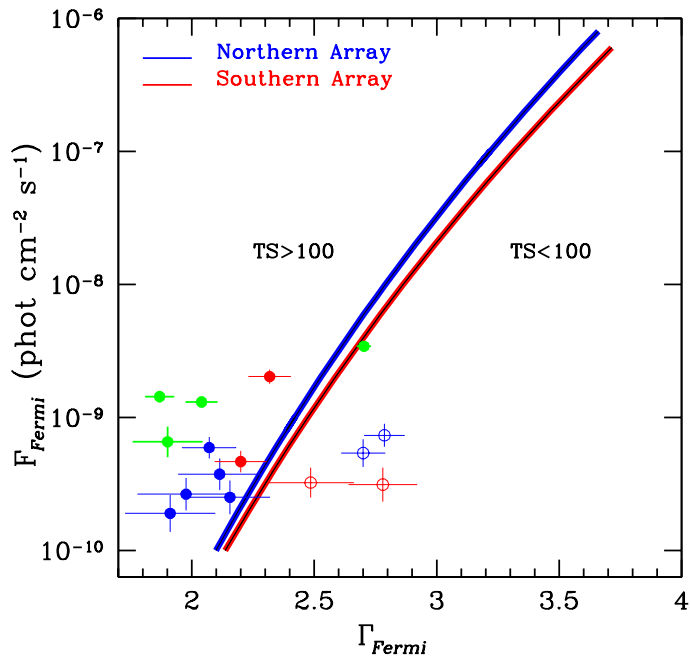}

\caption{Diagnostic space for sources with known photon index and flux in the 1$-100$~GeV band.
The curves  define the regions where a source can be detected by CTA assuming as a threshold TS$=$100.  The blue line refers to the Northern array, while the red one to the Southern array. The data points represent the \fermi\ radio galaxy
  sample, also distinguished by hemisphere. Filled points represent
 statistically significant detections, empty points represent sources with TS$<100$. Green points represent the radio galaxies already detected by the current Cherenkov Telescopes. NGC~1275 and NGC~6251 are not in figure: 
their curved \fermi\ spectra cannot be extrapolated with a power-law into the CTA band. The radio galaxy Centaurus~A falls exactly in the strip delimiting the detection from the non-detection regions.}
\label{ctsnr}
\end{figure*}

Figure~\ref{ctsnr} summarizes  the results: the input parameters, i.e. the 1--100~GeV flux and the power-law spectral slope, are reported on the x-axis and y-axis, respectively. The red (Southern hemisphere) and the blue (Northern hemisphere) curves connect the points of the grid  for which
a TS $=100$ is obtained for 50 hours of observation. 
The targets with a significance larger than {\bf $\sim10$} occupy the left part of the plot,  those  below the CTA sensitivity the right one. 
As a check of the reliability of our simulations, we  also plot  the previous studied MAGNs (red and blue circles) and the radio galaxies already detected by the Cherenkov telescopes (green points). They fall in the "correct" regions of the plot (compare with Table~\ref{tabpl}).   Cen~A falls exactly in the strip delimiting the detection 
from the non-detection regions, although an H.E.S.S spectrum is already available for this source \cite{aha09}.  This is clearly related to the hardening of the spectrum \cite{sa13} that our conservative simulation does not take into account.

We predict that  hard
sources  with $\Gamma_{\mathrm{Fermi}} \le 2.1$ can be easily revealed down to
1--100~GeV fluxes of the order of 10$^{-10}$~phot~cm$^{-2}$~s$^{-1}$ (as already
anticipated in Section 2, see also Figure 2), while  AGNs with
moderately steep slopes ($\Gamma_{\mathrm{Fermi}} \sim$ 2.4-2.8) require 1--100~GeV fluxes larger 
than 10$^{-9}$~phot~cm$^{-2}$~s$^{-1}$ to overcome the sensitivity threshold of the
Cherenkov arrays.  Sources with very steep spectra ($\Gamma_{\mathrm{Fermi}} >
2.8$) need 1--100~GeV fluxes as high as 10$^{-8}$--$10^{-7}$~phot~cm$^{-2}$~s$^{-1}$ to emerge in the TeV sky.

\section{Conclusions}
 \label{sec:con}
The aim of this work is to evaluate the impact of the next-generation
Cherenkov Telescope Array on TeV studies of Misaligned Active
Galactic Nuclei.
Up
to now, this class includes only 17 radio galaxies  detected at GeV
energies by \fermi\ (out of more than 3000 objects detected in
total) and 5 sources detected in the TeV band by Cherenkov telescopes
(out of about 60 radio-loud AGNs in the TeVCat online catalog).

In this study, we investigated the CTA detection prospects for MAGNs by simulating CTA observations for a sample of candidate sources observed by \fermi. The main results of our work can be summarized as follows.
\begin{itemize}
\item
We predict 8 new MAGNs at the $\sim10\sigma$ significance level, under the assumption of a straight extrapolation of the \fermi\ power-law in the CTA energy range.
\item
Assuming a power-law with an exponential cutoff as a more realistic
case, we still predict 5 (6)  new detections at a significance higher than $\sim10\sigma$ ($5\sigma$), for a high-energy cutoff of 500~GeV. 

\item
Additionally, our simulated data show that CTA will be able  to provide higher quality spectra with respect to current Cherenkov
facilities. We show this for the case of the newly TeV detected source PKS~0625$-$35.
\end{itemize}

Moreover, we generalize our conclusions by introducing a plot which
allows estimating whether a \fermi\ AGN will be detectable with the
CTA, given its flux and spectral slope in the 1--100~GeV energy range  and assuming a simple extrapolation of the  LAT power law spectrum into the TeV band. \\

We note that our estimates leave room for additional detections, since 
radio galaxies are variable sources, both in the \fermi\ band and
at Very High Energies. Additionally, a harder-when-brighter spectral
behavior has been observed in several \fermi\ AGNs, including radio
galaxies \cite{jan16}. This can further enhance even more the possibility of detecting
a larger number of sources during flaring states, thanks to the harder $\gamma$-ray
spectrum.

It should be also taken into account  that our sample is, by definition, biased towards 
sources with a high-energy SED peak in the \fermi\ band. This leaves
out  high-energy-peaked sources such as (or more extreme than) IC~310, 
which are faint in the \fermi\ band but should be detectable in the VHE regime.

 Finally, we point out that radio galaxies could have a more complex SED than expected. 
For example, there is evidence for a second spectral component in Cen~A 
that hardens the spectrum above 2 GeV (the slope changes from 2.7 a 2.1) \cite{sa13}. 
If this were a common feature in radio galaxies, MAGNs could be more easily revealed by CTA, opening at the same time new extremely appealing scenarios. \\

Our results indicate that long exposures are necessary, however, to study MAGNs that are in general steep and faint sources.
An extragalactic survey does not appear suitable for the exploration of this class of objects.
The CTA Key Science Projects, for example, include an extragalactic survey of 1/4th
of the sky in about 1000 hours \cite{dub13}. The area covered is of the order of
$\sim10^4~$deg$^2$. Given that the largest field-of-view
for the CTA will be $\sim8$\deg\  (for the SSTs), this implies an
effective observing time per pointing of the order of a few hours.
 An efficient strategy for radio galaxies seems to require long targeted campaigns with
effective observing times of the order of 50 hours.
This could be
achieved more easily, for example, by operating the CTA in subarrays. Because of the
relatively steep spectra of the sources, most of the emission will
fall in the energy band covered by the LSTs and MSTs. Therefore, it could be possible to observe this small sample of sources with these
telescopes as a subarray, while the SSTs gain exposure on an extreme
source of $\gamma$-rays at the highest energies.\\

In conclusion, we predict that the CTA will have a significant impact
in our understanding of MAGNs at TeV energies, with the likely
detection of additional sources. This would be an additional step
towards an understanding of this class of AGNs at TeV energies that relies
less on the properties of a single source, and more on the common
behavior.  Moreover,
obtaining better quality data on already detected sources would be
crucial in order to distinguish between different emission models to
explain their VHE emission.\\

\section*{Acknowledgements}
This paper has gone through internal review by the CTA Consortium. We would like to thank the two reviewers of the CTA Speaker's and Publication Office (SAPO) for their useful comments on the manuscript, and the anonymous referee for a helpful and constructive report.
We thank Valentina Fioretti for precious help in the installation of the software package {\it ctools}.
This research made use of {\it ctools}, a community-developed analysis package for Imaging Air Cherenkov Telescope data. {\it ctools} is based on GammaLib, a community-developed toolbox for the high-level analysis of astronomical gamma-ray data.

\section*{References}
\bibliography{mybibfile}

\begin{thebibliography}{10}
\expandafter\ifx\csname url\endcsname\relax
  \def\url#1{\texttt{#1}}\fi
\expandafter\ifx\csname urlprefix\endcsname\relax\def\urlprefix{URL }\fi
\expandafter\ifx\csname href\endcsname\relax
  \def\href#1#2{#2} \def\path#1{#1}\fi

\bibitem{miley80}
G.~{Miley}, \araa 18 (1980) 165.

\bibitem{bla}
R.~{Blandford}, W.~{East}, K.~{Nalewajko}, Y.~{Yuan}, J.~{Zrake},
  arXiv:1511.07515.

\bibitem{fo98}
G.~{Fossati}, L.~{Maraschi}, A.~{Celotti}, A.~{Comastri}, G.~{Ghisellini},
  \mnras 299 (1998) 433.

\bibitem{ma92}
L.~{Maraschi}, G.~{Ghisellini}, A.~{Celotti}, \apjl 397 (1992) L5.

\bibitem{sik}
M.~{Sikora}, M.~C. {Begelman}, M.~J. {Rees}, \apj 421 (1994) 153.

\bibitem{tav}
F.~{Tavecchio}, L.~{Maraschi}, R.~M. {Sambruna}, C.~M. {Urry}, \apjl 544 (2000)
  L23.

\bibitem{bott10}
M.~{B{\"o}ttcher}, Fermi Meets Jansky - AGN at Radio and Gamma-Rays", Eds.:
  Savolainen, T., Ros, E., Porcas, R. W., and Zensus, J. A, arXiv:1006.5048.

\bibitem{bott13}
M.~{B{\"o}ttcher}, A.~{Reimer}, K.~{Sweeney}, A.~{Prakash}, \apj 768 (2013) 54.

\bibitem{bar89}
P.~D. {Barthel}, \apj 336 (1989) 606.

\bibitem{urry95}
C.~M. {Urry}, P.~{Padovani}, \pasp 107 (1995) 803.

\bibitem{fan74}
B.~L. {Fanaroff}, J.~M. {Riley}, \mnras 167 (1974) 31.

\bibitem{p190}
P.~{Padovani}, C.~M. {Urry}, \apj 356 (1990) 75.

\bibitem{p290}
P.~{Padovani}, C.~M. {Urry}, \apj 368 (1991) 373.

\bibitem{p91}
C.~M. {Urry}, P.~{Padovani}, M.~{Stickel}, \apj 382 (1991) 501.

\bibitem{p92}
P.~{Padovani}, C.~M. {Urry}, \apj 387 (1992) 449.

\bibitem{bal15}
R.~D. {Baldi}, A.~{Capetti}, G.~{Giovannini}, \aap 576 (2015) A38.

\bibitem{mar02}
A.~P. {Marscher}, S.~G. {Jorstad}, J.-L. {G{\'o}mez}, M.~F. {Aller},
  H.~{Ter{\"a}sranta}, M.~L. {Lister}, A.~M. {Stirling}, \nat 417 (2002) 625.

\bibitem{cha09}
R.~{Chatterjee}, A.~P. {Marscher}, S.~G. {Jorstad}, A.~R. {Olmstead}, I.~M.
  {McHardy}, M.~F. {Aller}, H.~D. {Aller}, A.~{L{\"a}hteenm{\"a}ki},
  M.~{Tornikoski}, T.~{Hovatta}, K.~{Marshall}, H.~R. {Miller}, W.~T. {Ryle},
  B.~{Chicka}, A.~J. {Benker}, M.~C. {Bottorff}, D.~{Brokofsky}, J.~S.
  {Campbell}, T.~S. {Chonis}, C.~M. {Gaskell}, E.~R. {Gaynullina}, K.~N.
  {Grankin}, C.~H. {Hedrick}, M.~A. {Ibrahimov}, E.~S. {Klimek}, A.~K. {Kruse},
  S.~{Masatoshi}, T.~R. {Miller}, H.-J. {Pan}, E.~A. {Petersen}, B.~W.
  {Peterson}, Z.~{Shen}, D.~V. {Strel'nikov}, J.~{Tao}, A.~E. {Watkins},
  K.~{Wheeler}, \apj 704 (2009) 1689.

\bibitem{cha11}
R.~{Chatterjee}, A.~P. {Marscher}, S.~G. {Jorstad}, A.~{Markowitz},
  E.~{Rivers}, R.~E. {Rothschild}, I.~M. {McHardy}, M.~F. {Aller}, H.~D.
  {Aller}, A.~{L{\"a}hteenm{\"a}ki}, M.~{Tornikoski}, B.~{Harrison},
  I.~{Agudo}, J.~L. {G{\'o}mez}, B.~W. {Taylor}, M.~{Gurwell}, \apj 734 (2011)
  43.

\bibitem{chi1}
M.~{Chiaberge}, A.~{Celotti}, A.~{Capetti}, G.~{Ghisellini}, \aap 358 (2000)
  104.

\bibitem{abdo9a}
A.~A. {Abdo}, {et al.}, \apj 699 (2009) 31.

\bibitem{abdo9b}
A.~A. {Abdo}, {et al.}, \apj 707 (2009) 55.

\bibitem{abdo10a}
A.~A. {Abdo}, {et al.}, \apj 719 (2010) 1433.

\bibitem{mi}
G.~{Migliori}, P.~{Grandi}, E.~{Torresi}, C.~{Dermer}, J.~{Finke},
  A.~{Celotti}, R.~{Mukherjee}, M.~{Errando}, F.~{Gargano}, F.~{Giordano},
  M.~{Giroletti}, \aap 533 (2011) A72.

\bibitem{ghi15}
G.~{Ghisellini}, F.~{Tavecchio}, \mnras 448 (2015) 1060.

\bibitem{mar}
M.~{Georganopoulos}, D.~{Kazanas}, \apjl 594 (2003) L27.

\bibitem{gg2005}
G.~{Ghisellini}, F.~{Tavecchio}, M.~{Chiaberge}, A\&A 432 (2005) 401.

\bibitem{tave08}
F.~{Tavecchio}, G.~{Ghisellini}, MNRAS 385 (2008) L98.

\bibitem{tave14}
F.~{Tavecchio}, G.~{Ghisellini}, MNRAS 443 (2014) 1224.

\bibitem{giro2004}
M.~{Giroletti}, G.~{Giovannini}, L.~{Feretti}, W.~D. {Cotton}, P.~G. {Edwards},
  L.~{Lara}, A.~P. {Marscher}, J.~R. {Mattox}, B.~G. {Piner}, T.~{Venturi},
  \apj 600 (2004) 127.

\bibitem{ko2007}
Y.~Y. {Kovalev}, M.~L. {Lister}, D.~C. {Homan}, K.~I. {Kellermann}, \apjl 668
  (2007) L27.

\bibitem{na2014}
H.~{Nagai}, T.~{Haga}, G.~{Giovannini}, A.~{Doi}, M.~{Orienti}, F.~{D'Ammando},
  M.~{Kino}, M.~{Nakamura}, K.~{Asada}, K.~{Hada}, M.~{Giroletti}, \apj 785
  (2014) 53.

\bibitem{3LAC}
M.~{Ackermann}, {et al.}, \apj 810 (2015) 14.

\bibitem{2FHL}
M.~{Ackermann}, {et al.}, \apjs 222 (2016) 5.

\bibitem{ale2012}
J.~{Aleksi{\'c}}, {et al.}, A\&A 539 (2012) L2.

\bibitem{aha03}
F.~{Aharonian}, {et al.}, \aap 403 (2003) L1.

\bibitem{aha09}
F.~{Aharonian}, {et al.}, \apj 695 (2009) L40.

\bibitem{aleski09}
J.~{Aleksi{\'c}}, {et al.}, \apjl 723 (2010) L207.

\bibitem{dyr15}
M.~{Dyrda}, A.~{Wierzcholska}, O.~{Hervet}, R.~{Moderski}, M.~{Janiak},
  M.~{Ostrowski}, {\L}.~{Stawarz}, {for the H.~E.~S.~S.~Collaboration},
  Proceedings of the 34th International Cosmic Ray Conference (ICRC2015), The
  Hague, The Netherlands.\href {http://arxiv.org/abs/1509.06851}
  {\path{arXiv:1509.06851}}.

\bibitem{ach13}
B.~S. {Acharya}, M.~{Actis}, T.~{Aghajani}, G.~{Agnetta}, J.~{Aguilar},
  F.~{Aharonian}, M.~{Ajello}, A.~{Akhperjanian}, M.~{Alcubierre},
  J.~{Aleksi{\'c}}, et~al., Astroparticle Physics 43 (2013) 3.

\bibitem{abdo10b}
A.~A. {Abdo}, {et al.}, \apj 720 (2010) 912.

\bibitem{cas15}
C.~{Casadio}, J.~L. {G{\'o}mez}, P.~{Grandi}, S.~G. {Jorstad}, A.~P.
  {Marscher}, M.~L. {Lister}, Y.~Y. {Kovalev}, T.~{Savolainen}, A.~B.
  {Pushkarev}, \apj 808 (2015) 162.

\bibitem{gra16}
P.~{Grandi}, A.~{Capetti}, R.~D. {Baldi}, MNRAS 457 (2016) 2.

\bibitem{3FGL}
F.~{Acero}, {et al.}, \apjs 218 (2015) 23.

\bibitem{fornax}
M.~{Ackermann}, {et al.}, \apj 826 (2016) 1.

\bibitem{ale14ic}
J.~{Aleksi{\'c}}, {et al.}, \aap 563 (2014) A91.

\bibitem{aliu12}
E.~{Aliu}, { et al.}, \apj 746 (2012) 141.

\bibitem{nero10}
A.~{Neronov}, D.~{Semikoz}, I.~{Vovk}, \aap 519 (2010) L6.

\bibitem{ale14n}
J.~{Aleksi{\'c}}, {et al.}, \aap 564 (2014) A5.

\bibitem{aha06}
F.~{Aharonian}, {et al}, Science 314 (2006) 1424.

\bibitem{acc08}
V.~A. {Acciari}, {et al.}, \apj 679 (2008) 397.

\bibitem{abr12}
A.~{Abramowski}, F.~{Acero}, F.~{Aharonian}, A.~G. {Akhperjanian}, G.~{Anton},
  A.~{Balzer}, A.~{Barnacka}, U.~{Barres de Almeida}, Y.~{Becherini},
  J.~{Becker}, et~al., \apj 746 (2012) 151.

\bibitem{ale12l}
J.~{Aleksi{\'c}}, {et al.}, \aap 544 (2012) A96.

\bibitem{sa13}
N.~{Sahakyan}, R.~{Yang}, F.~A. {Aharonian}, F.~M. {Rieger}, \apj 770 (2013)
  L6.

\bibitem{fra14}
N.~{Fraija}, \mnras 441 (2014) 1209.

\bibitem{ri09}
F.~M. {Rieger}, F.~A. {Aharonian}, \aap 506 (2009) L41.

\bibitem{pe14}
M.~{Petropoulou}, E.~{Lefa}, S.~{Dimitrakoudis}, A.~{Mastichiadis}, \aap 562
  (2014) A12.

\bibitem{bro}
A.~M. {Brown}, C.~{Boehm}, J.~{Graham}, T.~{Lacroix}, P.~M. {Chadwick},
  J.~{Silk}, ArXiv e-prints\href {http://arxiv.org/abs/1603.05469}
  {\path{arXiv:1603.05469}}.

\bibitem{lobi}
A.~A. {Abdo}, {et al.}, Science 328 (2010) 725.

\bibitem{jan16}
M.~{Janiak}, M.~{Sikora}, R.~{Moderski}, \mnras 458 (2016) 2360.

\bibitem{3c111}
P.~{Grandi}, E.~{Torresi}, C.~{Stanghellini}, \apjl 751 (2012) L3.

\bibitem{kat}
J.~{Kataoka}, {\L}.~{Stawarz}, C.~C. {Cheung}, G.~{Tosti}, E.~{Cavazzuti},
  A.~{Celotti}, S.~{Nishino}, Y.~{Fukazawa}, D.~J. {Thompson}, W.~F.
  {McConville}, \apj 715 (2010) 554.

\bibitem{muk}
R.~{Mukherjee}, {VERITAS Collaboration}, The Astronomer's Telegram 9690.

\bibitem{kno16}
J.~{Kn{\"o}dlseder}, M.~{Mayer}, C.~{Deil}, J.-B. {Cayrou}, E.~{Owen},
  N.~{Kelley-Hoskins}, C.-C. {Lu}, R.~{Buehler}, F.~{Forest}, T.~{Louge},
  H.~{Siejkowski}, K.~{Kosack}, L.~{Gerard}, A.~{Schulz}, P.~{Martin},
  D.~{Sanchez}, S.~{Ohm}, T.~{Hassan}, S.~{Brau-Nogu{\'e}}, \aap 593 (2016) A1.

\bibitem{mat96}
J.~R. {Mattox}, D.~L. {Bertsch}, J.~{Chiang}, B.~L. {Dingus}, S.~W. {Digel},
  J.~A. {Esposito}, J.~M. {Fierro}, R.~C. {Hartman}, S.~D. {Hunter},
  G.~{Kanbach}, D.~A. {Kniffen}, Y.~C. {Lin}, D.~J. {Macomb}, H.~A.
  {Mayer-Hasselwander}, P.~F. {Michelson}, C.~{von Montigny}, R.~{Mukherjee},
  P.~L. {Nolan}, P.~V. {Ramanamurthy}, E.~{Schneid}, P.~{Sreekumar}, D.~J.
  {Thompson}, T.~D. {Willis}, \apj 461 (1996) 396.

\bibitem{nolan}
P.~L. {Nolan}, {et al.}, \apjs 199 (2012) 31.

\bibitem{dub13}
G.~{Dubus}, J.~L. {Contreras}, S.~{Funk}, Y.~{Gallant}, T.~{Hassan},
  J.~{Hinton}, Y.~{Inoue}, J.~{Kn{\"o}dlseder}, P.~{Martin}, N.~{Mirabal},
  M.~{de Naurois}, M.~{Renaud}, {CTA Consortium}, Astroparticle Physics 43
  (2013) 317.

\end{thebibliography}
\end{document}